\documentclass{aip-cp}
\pdfoutput=1
\usepackage[numbers]{natbib}
\usepackage{rotating}
\usepackage{graphicx}
\usepackage{subfigure}

\usepackage{aas_macros}

\begin{document}

\title{Statistical Measurement of the Gamma-ray Source-count Distribution as a Function of Energy}

\author[aff1,aff2]{H.-S. Zechlin\corref{cor1}}
\author[aff2,aff3]{A. Cuoco}
\author[aff1,aff2]{F. Donato}
\author[aff1,aff2]{N. Fornengo}
\author[aff1,aff2]{M. Regis}

\affil[aff1]{Dipartimento di Fisica, Universit\`a di Torino, via P. Giuria, 1, I-10125 Torino, Italy}
\affil[aff2]{Istituto Nazionale di Fisica Nucleare, Sezione di Torino, via P. Giuria, 1, I-10125 Torino, Italy}
\affil[aff3]{Institute for Theoretical Particle Physics and Cosmology (TTK), RWTH Aachen University, D-52056 Aachen, Germany}
\corresp[cor1]{Corresponding author: zechlin@to.infn.it}

\maketitle

\begin{abstract}
Photon counts statistics have recently been proven to provide a sensitive observable for characterizing gamma-ray source populations and for measuring the composition of the gamma-ray sky. In this work, we generalize the use of the standard 1-point probability distribution function (1pPDF) to decompose the high-latitude gamma-ray emission observed with \emph{Fermi}-LAT into: (i) point-source contributions, (ii) the Galactic foreground contribution, and (iii) a diffuse isotropic background contribution. We analyze gamma-ray data in five adjacent energy bands between 1 and 171\,GeV. We measure the source-count distribution $\mathrm{d}N/\mathrm{d}S$ as a function of energy, and demonstrate that our results extend current measurements from source catalogs to the regime of so far undetected sources. Our method improves the sensitivity for resolving point-source populations by about one order of magnitude in flux. The $\mathrm{d}N/\mathrm{d}S$ distribution as a function of flux is found to be compatible with a broken power law. We derive upper limits on further possible breaks as well as the angular power of unresolved sources. We discuss the composition of the gamma-ray sky and capabilities of the 1pPDF method.
\end{abstract}

\section{INTRODUCTION}
The extragalactic gamma-ray background (EGB; see Ref.~\cite{2015PhR...598....1F} for a recent review) comprises pivotal properties of the high-energy gamma-ray sky. Its composition is investigated by 
discriminating individual sources from diffuse components, utilizing gamma-ray measurements from spaceborne or ground-based instruments. Subtracting resolved point-source populations, which are characterized by their source-count distributions $\mathrm{d}N/\mathrm{d}S$ as a function of the integral flux $S$, leaves an unresolved background component that is nearly isotropic (IGRB; see Ref.~\cite{2015ApJ...799...86A}).

As opposed to conventional analyses methods of resolving point sources individually, statistical analysis techniques offer the possibility of decomposing gamma-ray sky maps into their constituents \cite{2011ApJ...738..181M,2015A&A...581A.126S,2016PhRvL.116e1103L,2016ApJS..225...18Z,2016ApJ...826L..31Z}. In Ref.~\cite{2016ApJS..225...18Z} and \cite{2016ApJ...826L..31Z}, we have demonstrated that the statistics of photon counts in single map pixels, i.e., the 1-point probability distribution function (1pPDF), can be used to accurately measure the total $\mathrm{d}N/\mathrm{d}S$ distribution of point sources and to decompose the gamma-ray sky. Given the statistical nature, the 1pPDF method has allowed us to measure the $\mathrm{d}N/\mathrm{d}S$ in the regime of so far unresolved faint point sources, i.e., down to fluxes a factor of $\sim$10 below nominal catalog detection thresholds. The analysis employed the six-year data provided by the \emph{Fermi} Large Area Telescope (\emph{Fermi}-LAT; see Ref.~\cite{2012ApJS..203....4A}) for high Galactic latitudes, in five energy bands between 1.04 and 171\,GeV. In the following, the results obtained in Ref.~\cite{2016ApJ...826L..31Z} are summarized briefly.

\section{1pPDF METHOD AND FERMI-LAT DATA}
The model of the gamma-ray sky comprised three components: (i) a population of gamma-ray point sources, described by $\mathrm{d}N/\mathrm{d}S$, which are distributed isotropically across the sky, (ii) a diffuse Galactic foreground component, and (iii) a diffuse component including all contributions indistinguishable from diffuse isotropic emission. The $\mathrm{d}N/\mathrm{d}S$ distribution was parameterized with a multiply broken power law (MBPL), including $N_\mathrm{b}$ free break positions and therefore $N_\mathrm{b}+1$ free power-law components connecting the breaks. The Galactic foreground (GF) emission was modeled with spatial and spectral templates, while the diffuse background emission was assumed to follow a power law spectrum. The normalization of the GF template, $A_\mathrm{gal}$, and the integral flux of the diffuse background component, $F_\mathrm{iso}$, were considered free fit parameters. We used the approach of probability generating functions developed in Ref.~\cite{2016ApJS..225...18Z} to compute the 1pPDF. The likelihood of the data is represented by a product over the probabilities of finding the number $k_p$ of measured counts in each individual pixel $p$,
therefore taking into account spatial variations of the GF component. Parameters were estimated from the profile likelihood (frequentist approach), as derived from MCMC sampling.

We used the first six years of \emph{Fermi}-LAT data (\texttt{P7REP\_CLEAN}) in five energy bands for high Galactic latitudes ($|b|\geq 30^\circ$): 1.04$-$1.99, 1.99$-$5.0, 5.0$-$10.4, 10.4$-$50.0, and 50$-$171\,GeV. Note that the last energy band was evaluated for Galactic latitudes $|b|\geq 10^\circ$ in order to provide sufficient counts statistics. Event selection and data processing were conducted as further explained in Ref.~\cite{2016ApJ...826L..31Z}.

\begin{figure*}[!t]
\begin{centering}
\subfigure[]{%
  \includegraphics[width=0.93\textwidth]{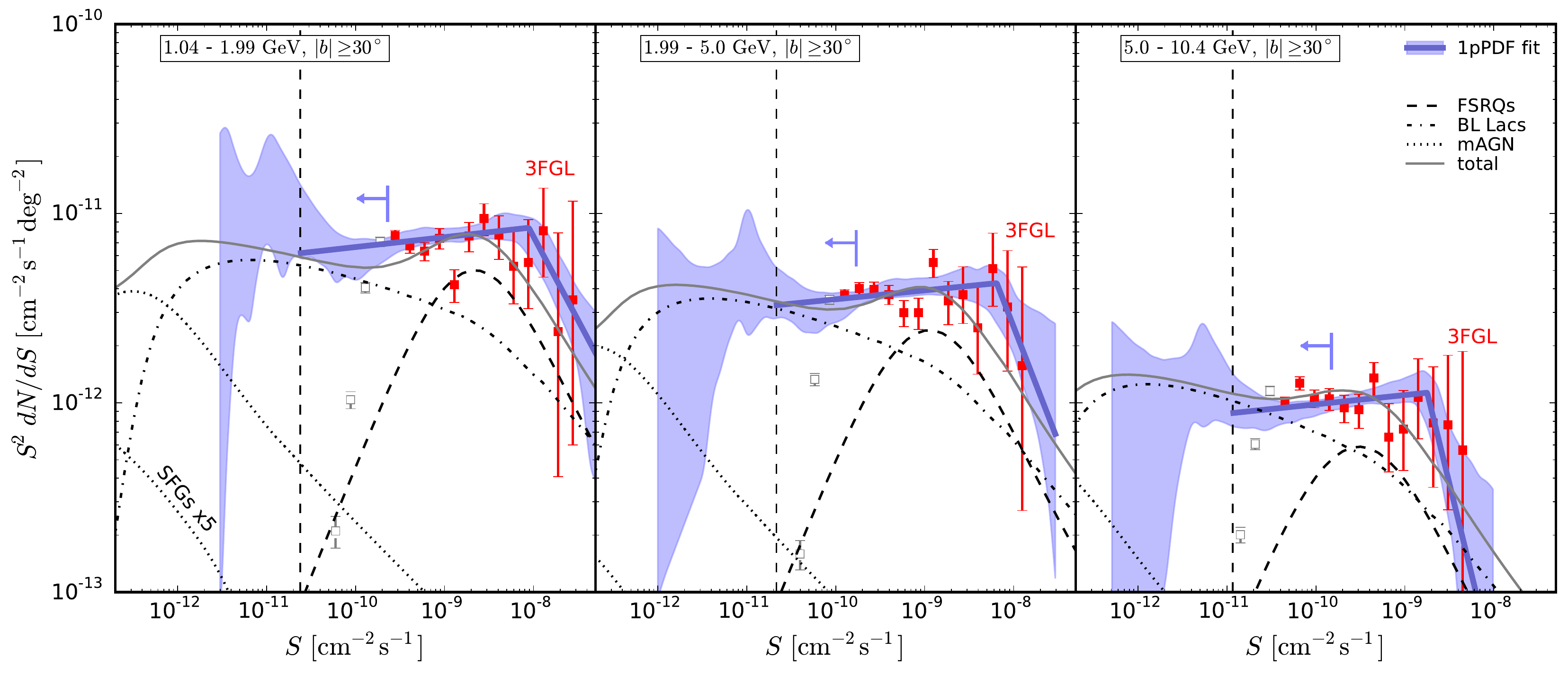}
  \label{sfig:dNdStop}
}\\
\subfigure[]{%
  \includegraphics[width=0.63\textwidth]{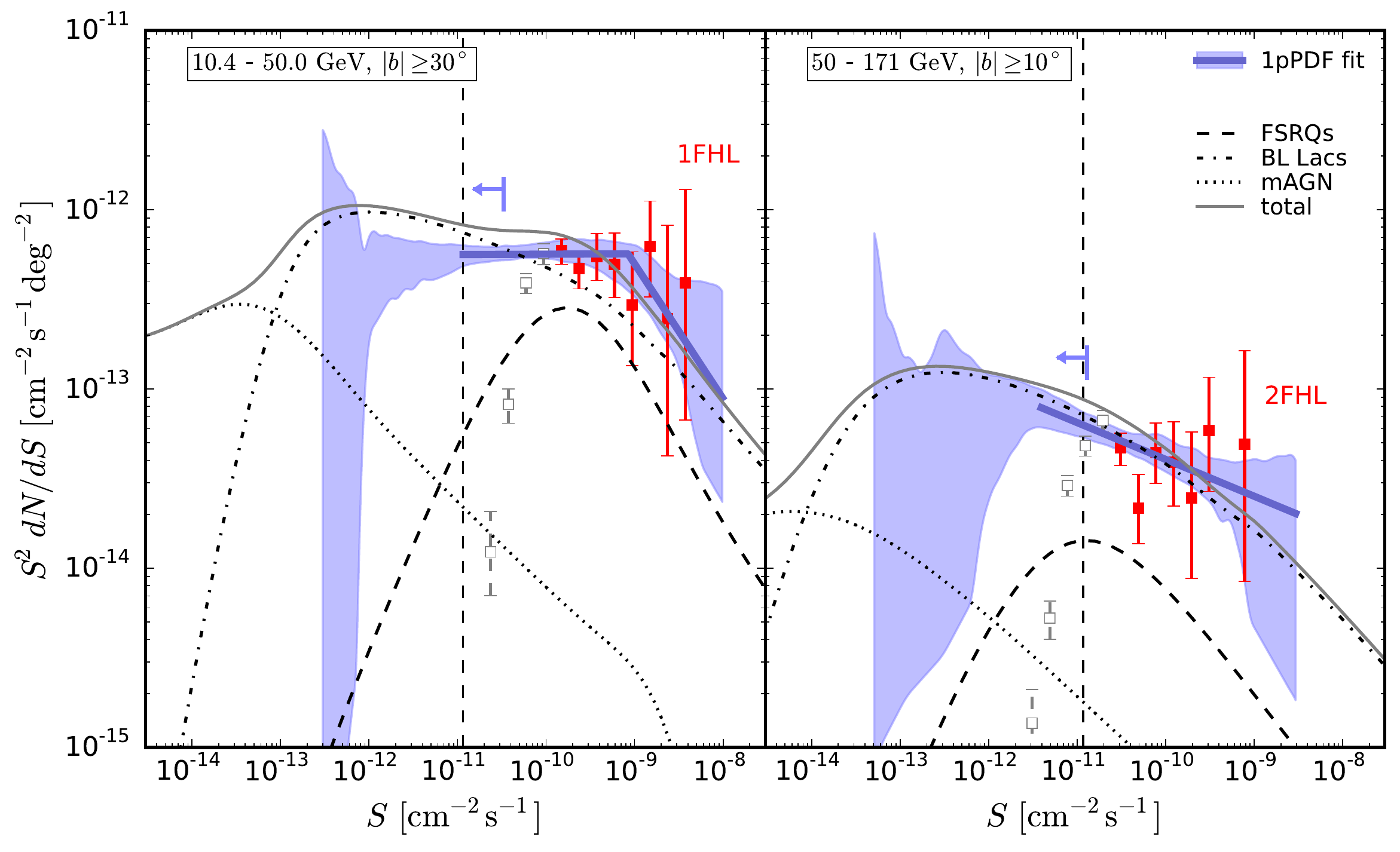}
  \label{sfig:dNdSbottom}
}
\caption{Differential source-count distributions $\mathrm{d}N/\mathrm{d}S$ obtained with the 1pPDF method, in five energy bands. The best-fits and the uncertainties at 68\% confidence level are depicted by the solid blue lines and the blue-shaded bands. The $\mathrm{d}N/\mathrm{d}S$ distributions derived from cataloged sources are shown by the red squares. The open gray squares depict $\mathrm{d}N/\mathrm{d}S$ points from sources below the nominal detection threshold; they thus cannot be used for comparison. The blue arrows denote upper limits (95\% confidence level) on a first or second break, respectively. The vertical dashed lines indicate the sensitivity estimates of the analysis. The dashed, dot-dashed, and dotted lines depict model predictions for flat spectrum radio quasars, BL~Lacertae objects, and misaligned active galactic nuclei; see Ref.~\cite{2016ApJ...826L..31Z} for details. \label{fig:dNdS}}
\end{centering}
\end{figure*}

\begin{figure}[t]
  \centerline{\includegraphics[width=0.6\textwidth]{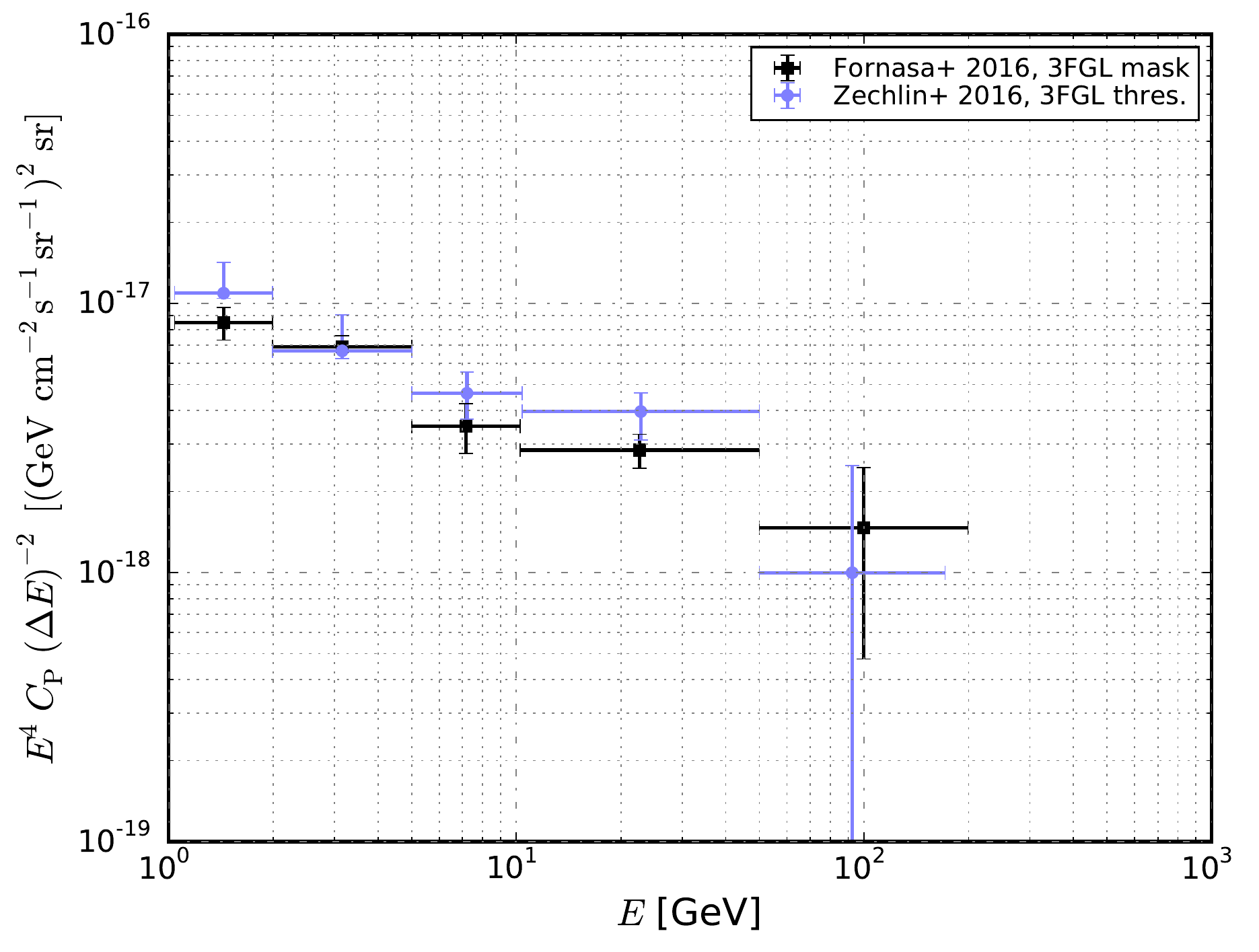}}
  \caption{Autocorrelation $C_\mathrm{P}$ as function of the energy $E$. The figure compares the $C_\mathrm{P}$ derived from the 1pPDF measurement of $\mathrm{d}N/\mathrm{d}S$ (blue circles) with the complementary anisotropy measurment of Ref.~\cite{2016arXiv160807289F} (black squares). See text for details. \label{fig:Cp}}
\end{figure}

\section{RESULTS}
\subsection{Source-count Distribution}
The techniques developed in Ref.~\cite{2016ApJS..225...18Z} were employed to fit our model of the gamma-ray sky to the data. We compared $\mathrm{d}N/\mathrm{d}S$ parameterizations by consecutively increasing the number of free breaks $N_\mathrm{b}$, in order to find the simplest parameterization required to fit the data properly.

The results are shown in Fig.~\ref{fig:dNdS}. We found that the data were described sufficiently well by broken power-law $\mathrm{d}N/\mathrm{d}S$ distributions, with a break at comparably high fluxes. The power-law indices below the first break are compatible with values between 1.95 and 2.0, except for the highest energy band that is compatible with a simple power-law fit with an index of $2.2^{+0.7}_{-0.3}$.
The best-fit solutions are depicted by the solid blue lines in Fig.~\ref{fig:dNdS}, which are shown only above the estimated sensitivity of the analysis. The blue shaded regions represent the corresponding uncertainty bands (68\% confidence level). Depending on the energy band, a higher number of breaks was assumed for their derivation, yielding robust and realistic estimates. The $\mathrm{d}N/\mathrm{d}S$ measurements are compared to the source counts of cataloged sources (derived from the catalogs mentioned in the figure, see Ref.~\cite{2016ApJ...826L..31Z}), well-matching our findings. It can be seen that the 1pPDF method extends the source-population sensitivity by almost one order of magnitude below the nominal catalog detection thresholds\footnote{Below the nominal threshold, the catalog suffers incompleteness.}.

Upper limits on a second (first, for the highest energy band) intrinsic break of $\mathrm{d}N/\mathrm{d}S$ are depicted by the blue arrows in Fig.~\ref{fig:dNdS}.

\subsection{Anisotropies}
A complementary measure of unresolved point sources is provided by anisotropy measurements (see, e.g., Ref.~\cite{2016arXiv160807289F}). In particular, the source-count distribution is related to the autocorrelation angular power spectrum $C_\mathrm{P}$ by
\begin{equation}\label{eq:Cp}
C_\mathrm{P}(S_{\rm th}) =  \int_0^{S_{\rm th}} S^2
\frac{\mathrm{d}N}{\mathrm{d}S} \mathrm{d}S,
\end{equation}
where $S_{\rm th}$ is the flux threshold of individually resolved point sources. Figure~\ref{fig:Cp} compares the autocorrelation derived from our $\mathrm{d}N/\mathrm{d}S$ measurements with the complementary anisotropy analysis in Ref.~\cite{2016arXiv160807289F}. Since the analysis of the afore-mentioned reference refers to the 3FGL catalog for source-masking, we defined the anisotropy below the effective detection threshold of the 3FGL catalog, $C_\mathrm{P}(S^\mathrm{3FGL}_\mathrm{th})$, by means of Eq.~\ref{eq:Cp}, such that $C_\mathrm{P}(S^\mathrm{3FGL}_\mathrm{th}) \approx C_\mathrm{P}(S^{<1}_\mathrm{th}) - C^\mathrm{cat}_\mathrm{P}(S^{<1}_\mathrm{th})$, where $C^\mathrm{cat}_\mathrm{P}$ denotes the anisotropy contributed by cataloged (i.e., resolved) sources only, and $S^{<1}_\mathrm{th}$ approximates the nominal catalog detection threshold of the 3FGL. Figure~\ref{fig:Cp} demonstrates that the two analyses agree within uncertainties.

\subsection{Composition of the Gamma-ray Sky}
The integral fluxes of the three considered components, i.e., the point-source flux $F_\mathrm{ps}$, the GF template flux $F_\mathrm{gal}$, and the flux of the isotropic diffuse background component $F_\mathrm{iso}$, provide the decomposition of the gamma-ray sky. In Tab.~\ref{tab:sky}, the flux contributions are compared to the total flux of the high-latitude gamma-ray sky. The contribution from point sources can also be expressed as fractional contribution to the extragalactic gamma-ray background (EGB), $F_\mathrm{EGB}$, as measured in Ref.~\cite{2015ApJ...799...86A}. The resulting contributions $F_\mathrm{ps}/F_\mathrm{EGB}$ in each energy band are $0.83^{+0.07}_{-0.13}$, $0.79^{+0.04}_{-0.16}$, $0.66^{+0.20}_{-0.07}$, $0.66^{+0.28}_{-0.05}$, and $0.81^{+0.52}_{-0.19}$, respectively.

\begin{table}[t]
\caption{Composition of the Gamma-ray Sky. The quantities $q_\mathrm{ps}$, $q_\mathrm{gal}$, and $q_\mathrm{iso}$ are the ratios of the integral flux components ($F_\mathrm{ps}$, $F_\mathrm{gal}$, $F_\mathrm{iso}$) and the total map flux $F_\mathrm{tot}$. Parentheses denote symmetric errors on the preceding digit. The parameter $F_\mathrm{tot}$ is given in units of $\mathrm{cm}^{-2}\,\mathrm{s}^{-1}\,\mathrm{sr}^{-1}$.}
\label{tab:sky}
\tabcolsep7pt
\begin{tabular}{lccccc}
\hline
\tch{1}{c}{b}{Parameter}  &  \tch{1}{c}{b}{$1.04-1.99\,\mathrm{GeV}$}  &  \tch{1}{c}{b}{$1.99-5.0\,\mathrm{GeV}$} &  \tch{1}{c}{b}{$5.0-10.4\,\mathrm{GeV}$} &  \tch{1}{c}{b}{$10.4-50.0\,\mathrm{GeV}$} 
  &  \tch{1}{c}{b}{$50-171\,\mathrm{GeV}$} \\
\hline
$q_\mathrm{ps}$ & $0.27^{+0.02}_{-0.04}$ &  $0.27^{+0.02}_{-0.06}$
& $0.24^{+0.08}_{-0.03}$ &  $0.27^{+0.11}_{-0.03}$ &  $0.29^{+0.19}_{-0.08}$ \\[0.3em]
$q_\mathrm{gal}$ & $0.714^{+0.003}_{-0.005}$ &  $0.708^{+0.005}_{-0.006}$
& $0.598^{+0.007}_{-0.01}$ &  $0.494^{+0.008}_{-0.013}$ &  $0.49^{+0.01}_{-0.02}$ \\[0.3em]
$q_\mathrm{iso}$ & $0.02^{+0.04}_{-0.01}$ &  $0.012^{+0.061}_{-0.001}$
& $0.16^{+0.03}_{-0.07}$ &  $0.23^{+0.04}_{-0.15}$ &  $0.22^{+0.10}_{-0.17}$ \\[0.3em]
$F_\mathrm{tot}$ & $9.17(1) \times 10^{-7}$ &  $4.573(9) \times 10^{-7}$
& $1.103(3) \times 10^{-7}$ & $5.27(2) \times 10^{-8}$  &  $5.67(6) \times 10^{-9}$ \\[0.3em]
\hline
\end{tabular}
\end{table}

\section{ACKNOWLEDGMENTS}
This work is supported by the research grant {\sl Theoretical Astroparticle Physics} number 2012CPPYP7 under the program PRIN 2012 funded by the Ministero dell'Istruzione, Universit\`a e della Ricerca (MIUR), by the research grants {\sl TAsP (Theoretical Astroparticle Physics)} and {\sl Fermi} funded by the Istituto Nazionale di Fisica Nucleare (INFN), and by the {\sl Strategic Research Grant: Origin and Detection of Galactic and Extragalactic Cosmic Rays} as well as {\sl Excellent Young PI Grant: The Particle Dark-matter Quest in the Extragalactic Sky} funded by Torino University and Compagnia di San Paolo.

\nocite{*}
\bibliographystyle{aipnum-cp}%
\bibliography{zechlin_gamma2016}%

\end{document}